\documentclass{iaus}
\usepackage{graphicx}

\title[Horizontal Branch Stars in Globular Clusters]{Horizontal Branch
A- and B-type Stars in Globular Clusters} 

\author[S. Moehler]%
{S. Moehler}
\affiliation{Institut f\"ur theoretische Physik und Astrophysik,
Christian-Albrechts-Universit\"at zu Kiel, Olshausenstra\ss e 40, D
24118 Kiel, Germany; email: moehler@astrophysik.uni-kiel.de}

\pubyear{2004}
\volume{224}  
\pagerange{119--126}
\date{?? and in revised form ??}
\jname{The A-Star Puzzle}
\editors{J. Zverko, W.W. Weiss, J. Ziznovsky \& S.J. Adelman, eds.}

\begin{document}

\maketitle

\begin{abstract}
Globular clusters offer ideal laboratories to test the predictions of
stellar evolution. When doing so with spectroscopic analyses during
the 1990s, however, the parameters we derived for hot horizontal
branch stars deviated systematically from theoretical predictions.
The parameters of cooler, A-type horizontal branch stars, on the other hand,
were consistent with evolutionary theories.  In 1999, two groups
independently suggested that diffusion effects might cause these
deviations, which we verified subsequently. I will discuss these
observations and analyses and their consequences for interpreting
observations of hot horizontal branch stars.
\end{abstract}

\firstsection 
\section{Historical Background\label{history}}

Globular clusters are densely packed, gravitationally bound systems of
several thousands to about one million stars. The dimensions of the
globular clusters are small compared to their distance from us: half
of the light is generally emitted within a radius of less than 10~pc,
whereas the closest globular cluster has a distance of 2~kpc and 90\%
lie more than 5~kpc away. We can thus safely assume that all stars
within a globular cluster lie at the same distance from us. With ages
in the order of $10^{10}$~years globular clusters are among the oldest
objects in our Galaxy. As they formed stars only once in the beginning
and the duration of that star formation episode is short compared to
the current age of the globular clusters the stars within one globular
cluster are essentially coeval. In addition all stars within one
globular cluster (with few exceptions) show the same initial abundance
pattern (which may differ from one cluster to another).  Globular
clusters are thus the closest approximation to a physicist's
laboratory in astronomy.

The horizontal branch, which is the topic of this article, was
discovered by \cite{tebr27}, when he used Shapley's
\cite[(1915)]{shap1915} data on M~3 and other clusters to plot
magnitude versus colour (replacing luminosity and spectral type in the
Hertzsprung-Russell diagram) and thus produced the first
colour-magnitude diagrams ({\sc ``Farbenhelligkeitsdiagramme''}). In
these colour-magnitude diagrams (CMD's) ten Bruggencate noted the
presence of a red giant branch that became bluer towards fainter
magnitudes, in agreement with \cite{shap15}.  In addition, however,
he saw a horizontal branch ({\sc ``horizontaler Ast''}) that parted
from the red giant branch and extended far to the blue at constant
brightness.
As more CMD's of globular clusters were obtained it became obvious
that the relative numbers of red and blue horizontal branch stars
(i.e.\ the horizontal branch morphology) varied quite considerably
between individual clusters, with some clusters showing extensions of
the blue horizontal branch (so-called ``blue tails'') towards bluer
colours and fainter visual magnitudes, i.e. towards hotter
temperatures\footnote{The change in slope of the horizontal branch
towards higher temperatures is caused by the decreasing sensitivity of
$B-V$ to temperature on one hand and by the increasing bolometric
correction for hotter stars (i.e. the maximum of stellar flux is
radiated at ever shorter wavelengths for increasing temperatures,
making stars fainter at $V$) on the other hand.}.
\cite{sawa60}  noted a correlation between the metal
abundance and the horizontal branch morphology seen in globular cluster CMD's:
the horizontal branch ({\bf HB}) became bluer with decreasing metallicity. 

About 25 years after the discovery of the horizontal branch
\cite{hosc55} were the first to identify horizontal branch stars with
post-red giant branch stars that burn helium in the central regions of
their cores. \cite{faul66} managed for the first time to compute zero
age horizontal branch ({\bf ZAHB}) models that qualitatively
reproduced the observed trend of HB morphology with metallicity
without taking into account any mass loss but assuming a rather high
helium abundance of Y = 0.35. \cite{ibro70}, however, found that they
could {\it ``\ldots account for the observed spread in colour along the
horizontal branch by accepting that there is also a spread in stellar
mass along this branch, bluer stars being less massive (on the
average) and less luminous than redder stars.''  }
Comparing HB models to observed globular cluster CMD's 
\cite{rood73} found that an HB that 
{\it
``\ldots is made up of stars with the same core mass and slightly varying total
mass, produces theoretical c-m diagrams very similar to those observed. \ldots
A mass loss of perhaps 0.2~M$_\odot$ with a random dispersion of several
hundredths of a solar mass is required somewhere along the giant branch.''
}
The assumption of mass loss on the red giant branch
diminished the need for very high helium
abundances.

Thus our current understanding sees horizontal branch stars as stars
that burn helium in a core of about 0.5~M$_\odot$\ and hydrogen in a
shell and evolve to the asymptotic giant branch, when the helium in
the core is exhausted \cite[(for a review
on HB evolution see Sweigart 1994)]{swei94}. 
The more massive the hydrogen envelope is the cooler is the
resulting star at a given metallicity\footnote{Due to the higher opacities in
their envelopes metal-rich HB stars are cooler than metal-poor ones
with the same envelope mass.}. The masses of the hydrogen
envelopes vary from $\le$0.02~M$_\odot$\ at the hot end of the horizontal
branch (about 30,000~K) to 0.3--0.4~M$_\odot$\ for the cool
HB stars at about 4000--5000~K \cite[(depending on metallicity, e.g. Dorman
et al. 1993)]{doro93}. The stable red and blue HB stars are separated
by the variable RR Lyrae range at about 6500--7500~K. This article
deals with blue HB stars, which at effective temperatures of about
8,000~K to 20,000~K show spectra rather similar (at moderate
resolution) to main sequence stars of spectral types A and B and are
therefore called HBA and HBB stars. In the field of the Milky Way such 
stars are often denominated by FHB (field HB star) and used as tracers 
for halo structure.

\section{Atmospheric parameters\label{other}}
\begin{figure}
\vspace*{10cm}
\includegraphics{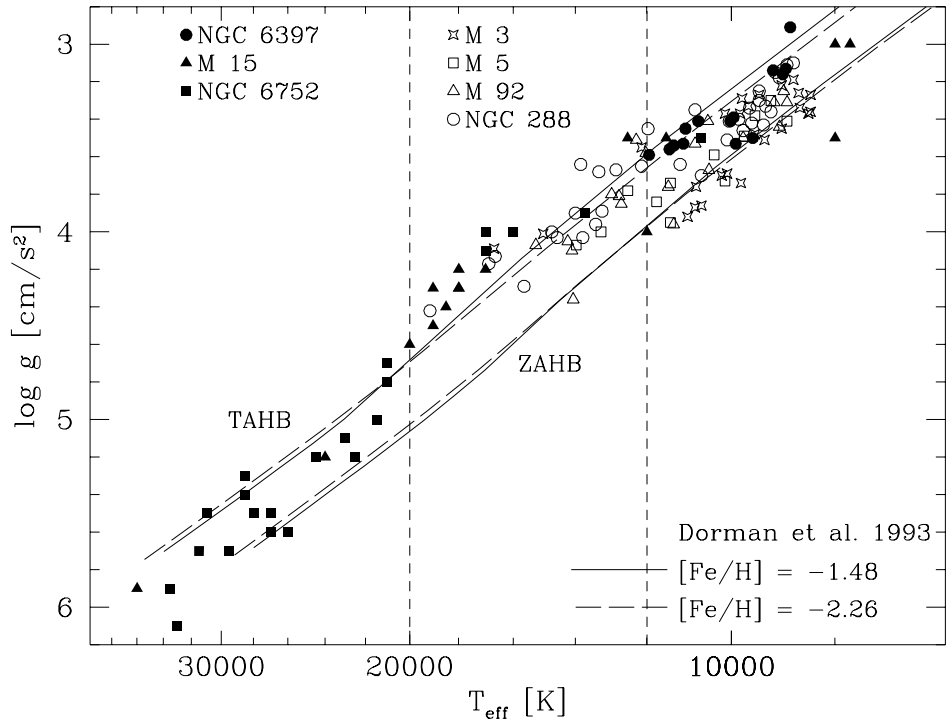}
\caption[]{The results of 
\cite[Crocker \etal\ (1988, M~3, M~5, M~92, NGC~288)]{crro88}, 
\cite[de Boer \etal\ (1995, NGC~6397)]{dbsc95},
\cite[Moehler \etal\ (1995, 1997a, M~15)]{mohe95,mohe97a}, and
\cite[Moehler \etal\ (1997b, NGC~6752)]{mohe97b} compared
to evolutionary tracks from \cite{doro93}. ZAHB and TAHB stand for
zero age and terminal age HB (see text for details).  The short-dashed
lines mark the regions of low log\,g\ (see text for details). \\
\label{ag_plottga1}}
\end{figure}

Already early studies of HBA and HBB stars in globular clusters showed 
discrepancies between observational results and theoretical expectations:
\cite{grdo66} mentioned that the comparison of $(c_1)_0$ vs.
$(b-y)_0$ for 50 blue HB stars in NGC~6397 to models from 
\cite{miha66} indicated low surface gravities and a mean mass of
0.3M$_\odot$\ (0.4M$_\odot$ ) for solar (negligible) helium abundance,
assuming $(m-M)_0 =$ 12.0 and E$_{\rm B-V}$\ = 0.16.  Later
spectroscopic analyses of HB stars (see cited papers for details) in
globular clusters reproduced this effect (cf. Fig.~\ref{ag_plottga1}):
\cite[Crocker \etal\ (1988, M~3, M~5, M~15, M~92,
NGC~288)]{crro88}, \cite[de Boer \etal\ (1995, NGC~6397)]{dbsc95},
\cite[Moehler \etal\ (1995, 1997a, M~15)]{mohe95,mohe97a},
\cite[Moehler \etal\ (1997b, NGC~6752)]{mohe97b}.

The zero age HB (ZAHB) in Fig.~\ref{ag_plottga1} marks the position
where the HB stars have settled down and started to quietly burn
helium in their cores.  The terminal age HB (TAHB) is defined by
helium exhaustion in the core of the HB star ($Y_C < 0.0001$). For
temperatures between 12,000~K and 20,000~K the observed positions in
the (log\,g, T$_{\rm eff}$)-diagram fall mostly above the ZAHB and in
some cases even above the TAHB. This agrees with the finding of
\cite{sake97} that field HBB stars show a larger scatter away from the
ZAHB in T$_{\rm eff}$, log\,g\ than the hotter subdwarf B stars 
with T$_{\rm eff} >$ 20,000~K.
Knowing the atmospheric parameters of the stars and the distances to the
globular clusters we can determine masses for the stars
\cite[(cf. Moehler \etal\ 1995, 1997b, de Boer \etal\ 
1995)]{mohe95,mohe97b,dbsc95}. While the stars in M~3, M~5, and
NGC~6752 have mean masses consistent with the canonical values, the
hot HB stars in all other clusters show masses that are significantly
lower than predicted by canonical HB evolution -- even for
temperatures cooler than 12,000~K.

Also some UV observations suggest discrepancies between theoretical
expectations and observational results: The IUE (International
Ultraviolet Explorer) and HUT (Hopkins Ultraviolet Telescope) spectra
of M~79 \cite[(Altner \& Matilsky 1993, Dixon \etal\
1996)]{alma93,dida96} suggest lower than expected gravities and higher
than expected metallicities for hot HB stars
\cite[(but see Vink \etal\ 1999, who do not need low surface gravities to fit 
the HUT data)]{vihe99}. \cite{hich96} find from Ultraviolet Imaging
Telescope (UIT) photometry of M~79 that stars bluer than
$m_{152}-m_{249}$ = $-$0.2 lie above the ZAHB, whereas cooler stars
scatter around the ZAHB. UIT data of M~13 \cite[(Parise \etal\
1998)]{pabo98} find a lack of stars close to the ZAHB at a colour
(temperature) range similar to the low log\,g\ range shown in
Fig.~\ref{ag_plottga1}.
\cite{lasw96} on the other hand find good agreement between UIT
photometry of blue stars in NGC~6752 and a standard ZAHB.

\section{Atmospheric abundances\label{sec-diff}}
It has been realized early on that the blue HB and blue tail stars in
globular clusters show weaker helium lines than field main sequence B stars
of similar temperatures: \cite[Searle \& Rodgers (1966, NGC~6397)]{sero66}; 
\cite[Greenstein \& M\"unch (1966, M~5, M~13, M~92)]{grmu66}; 
\cite[Sargent (1967, M~13, M~15, M~92)]{sarg67}. 
\cite{grtr67} already suggested
diffusion to explain this He deficiency.
\cite{miva83} performed the first theoretical study of
diffusion effects in hot horizontal branch stars. Using
the evolutionary tracks of \cite{swgr76} they found for the metal-poor
models that {\it ``in most of each envelope, the radiative
acceleration on all elements {\rm (i.e. C, N, O, Ca, Fe)} is much
larger than gravity which is not the case in main-sequence stars.''}
The elements are thus pushed towards the surface of the
star. Turbulence affects the different elements to varying extent, but
generally reduces the overabundances\footnote{\cite{mich82} and
\cite{chmi88} showed that meridional circulation can prevent
gravitational settling and that the limiting rotational velocity
decreases with decreasing log\,g. \cite{beco00} note that two of the
HB stars hotter than 10,000~K show higher rotational velocities and
much smaller abundance deviations than other stars of similar
temperature.}. Models without turbulence and/or mass loss (which may
reduce the effects of diffusion) predict stronger He depletions than
observed. A weak stellar wind could alleviate this discrepancy
\cite[(Heber 1986, Michaud \etal\ 1989, Fontaine \& Chayer 1997, and
Unglaub \& Bues 1998 discuss this effect, albeit for hotter
stars)]{hebe86,mibe89,foch97,unbu98}.
The extent of the predicted abundance variations varies with effective
temperature, from none for HB stars cooler than about $5800 \pm 500$K
(due to the very long diffusion timescales) to 2 -- 4 dex in the
hotter stars (the hottest model has T$_{\rm eff}$\ = 20,700~K) and
also depends on the element considered. 

Observations of blue HB and blue tail stars in globular clusters support the
idea of diffusion being active above a certain temperature:
Abundance analyses of blue HB stars cooler than 11,000~K to 12,000~K
in general show no deviations from the globular cluster abundances
derived from red giants, while for hotter stars departures from the
general globular cluster abundances are found, e.g.\ iron enrichment
to solar or even super-solar values and strong helium depletion 
(see Moehler 2001 for references and more details).
This agrees with the finding of \cite{alma93} and \cite{vihe99}
that solar metallicity model atmospheres are required 
to fit the UV spectra of M~79. 

All this evidence supports the suggestion of \cite{grca99} that the
onset of diffusion in stellar atmospheres may play a r\^ole in
explaining the jump along the HB towards brighter $u$ magnitudes at
effective temperatures of about 11,500~K. This jump in $u, u-y$ is
seen in all CMD's of globular clusters that have Str\"omgren
photometry of sufficient quality. The effective temperature of the
jump is roughly the same for all clusters, irrespective of
metallicity, central density, concentration or mixing evidence, and
coincides roughly with the onset of the ``low log\,g problem'' seen in
Fig.~\ref{ag_plottga1} at T$_{\rm eff}$\ $\approx$11,000~K to
12,000~K. This in turn coincides with the region where surface
convection zones due to hydrogen and He\,{\sc i} ionization disappear
in HB stars \cite[(Sweigart 2002)]{swei01}.

Radiative levitation of heavy elements decreases the far-UV flux and by
backwarming increases the flux in $u$. \cite{grca99} show that the use of
metal-rich atmospheres ([Fe/H] = $+0.5$ for scaled-solar ATLAS9 Kurucz
model atmospheres with $\log \epsilon_{Fe,\odot} = 7.60$) improves the
agreement between observed data and theoretical ZAHB in the $u, u-y$-CMD at
effective temperatures between 11,500~K and 20,000~K, 
but it worsens the agreement between theory and
observation for hotter stars in the Str\"omgren CMD of NGC~6752 (see their
Fig.~8). Thus diffusion may either not be as important in the hotter stars
or the effects may be diminished by a weak stellar wind.
The gap at $(B-V)_0 \approx 0$ discussed by 
\cite{calo99} 
is not directly related to the $u$-jump as it corresponds to an
effective temperature of about 9000~K and is also not seen in every
cluster (which would be expected if it were due to an atmospheric
phenomenon).  The gap at T$_{\rm eff}$\ $\approx$ 13,000~K seen in the
$c_1, b-y$ diagram of field horizontal branch stars \cite[(Newell
1973, Newell \& Graham 1976)]{newe73,negr76} may be related to the
$u$-jump as the $c_1$ index contains $u$.

The abundance distribution within a stellar atmosphere influences the
temperature stratification and thereby the line profiles and the flux
distribution of the emergent spectrum. A deviation in atmospheric
abundances of HB stars from the cluster metallicity due to diffusion
would thus affect their line profiles and flux distribution. Model
atmospheres calculated for the cluster metallicity may then yield
wrong results for effective temperatures and surface gravities when
compared to observed spectra of HB stars. This effect could explain at
least part of the observed discrepancies (see Sect.~\ref{hb_final} for
more details). Self-consistent model atmospheres taking into account
the effects of gravitational settling and radiative levitation are,
however, quite costly in CPU time and have started to appear only
quite recently for hot stars
\cite[(Dreizler \& Wolff 1999, Hui-Bon-Hoa \etal\
2000)]{drwo99,hule00}. 

\section{Rotational velocities\label{rotation}}

\cite[Peterson (1983, 1985a, 1985b)]{pete83,pete85a,pete85b}
found from high-resolution spectroscopic studies that clusters with
bluer HB morphologies show higher rotation velocities among their HB
stars.  However, the analysis of \cite{pero95} shows that while the
stars in M~13 (which has a long blue tail) rotate on average faster
than those in M~3 (which has only a short blue HB), the stars in
NGC~288 and M~13 show {\em slower} rotation velocities at {\em higher}
temperatures. These results are consistent with those reported by
\cite{bedj00}, who determined rotational velocities for stars as hot
as 19,000~K in M~13. They found that stars hotter than about 11,000~K
have significantly lower rotational velocities than cooler stars and
that the change in mean rotational velocity may coincide with the gap
seen along the blue HB of M~13 \cite[(see Ferraro \etal\ 1998 for an
extensive discussion of gaps)]{fepa98}.  Also the results of
\cite[Cohen \& McCarthy (1997, M~92)]{comc97} and \cite[Behr \etal\
(2000b, M15)]{beco00} show that HB stars cooler than $\approx$11,000~K
to 12,000~K in general rotate faster than hotter stars.

The studies by \cite{behr03a} and \cite[Recio \etal\ (2002,
2004)]{repi02,repi04}, which both consider several clusters, confirm
the trends mentioned above. Below the diffusion threshold about 20\%
to 30\% of the blue HB stars show rotation velocities of $v \sin i
\approx 20 \ldots 30$ km/s, whereas there are no fast rotators among
the hotter stars. From observations of field HB stars \cite{behr03b}
and \cite{cala03} note that red and cool blue HB stars show similar
distributions of rotational velocities (after accounting for the
larger radii of the red HB stars), whereas field RR Lyrae stars show no
evidence for rotation.

\cite{sipi00} study theoretical models for the rotation of HB stars and find
that the observed rotation of cool blue HB stars in M~13 can be
explained if their red giant precursors have rapidly rotating cores
and differential rotation in their convective envelopes and if angular
momentum is redistributed from the rapidly rotating core to the
envelope on the horizontal branch. If, however, turn-off stars rotate
with less than 4~km/s, a rapidly rotating core in the main-sequence
stars (violating helioseismological results for the Sun) or an
additional source of angular momentum on the red giant branch \cite[(e.g. mass
transfer in close binaries or capture of a planet as described
by Soker \& Harpaz 2000)]{soha00} are required to explain the rotation
of blue HB stars.
\cite{tach04} speculate that such differential rotation could be
understood within the framework of internal gravitational waves. The
change in rotation rates towards higher temperatures is not predicted
by the models of \cite{sipi00} but could be understood as a result of
gravitational settling of helium, which creates a mean molecular weight
gradient, that then inhibits angular momentum transport to the surface of the
star. \cite{swei01} suggests that the weak stellar wind invoked to
reconcile observed abundances in hot HB stars with
diffusion calculations (cf. Sect.~\ref{sec-diff}) could also carry
away angular momentum from the surface layers and thus reduce the
rotational velocities of these stars.

\cite{soha00} argue that the distribution of rotational velocities along
the HB can be explained by spin-up of the progenitors due to
interaction with low-mass companions, predominantly gas-giant planets,
in some cases also brown dwarfs or low-mass main-sequence stars
(esp. for the very hot HB stars). The slower rotation of the
hotter stars in their scenario is explained by mass loss {\em on} the
HB, which is accompanied by efficient angular momentum loss.  This
scenario, however, does not explain the sudden change in rotational
velocities and the coincidence of this change with the onset of
radiative levitation.

\section{Where do we stand?\label{hb_final}}

Analysis of a larger sample of hot HB stars in NGC~6752
\cite[(Moehler \etal\ 2000)]{mosw00} showed that the use of model
atmospheres with solar or super-solar abundances removes much of the
deviation from canonical tracks in effective temperature, surface
gravity and mass for hot HB stars discussed in
Sect.~\ref{other}. However, some discrepancies remain, indicating that
the low log\,g, low mass problem cannot be completely solved by
scaled-solar metal-rich atmospheres \cite[(which {\em do} reproduce
the $u$-jump reported by Grundahl \etal\ 1999)]{grca99}.  As
\cite{miva83} noted diffusion will not necessarily enhance all heavy
elements by the same amount and the effects of diffusion vary with
effective temperature. Elements that were originally very rare may be
enhanced even stronger than iron. The
question of whether diffusion can fully explain the ``low gravity''
problem cannot be answered without detailed abundance analyses to
determine the actual abundances and subsequent analyses using model
atmospheres with non-scaled solar abundances
\cite[(like ATLAS12, Kurucz 1992)]{kuru92}. Additional caution is
also recommended by the results of \cite{mola03} on M~13, where even
the use of the metal-rich model atmospheres does not eliminate the
problem of the low masses or the low gravities for effective
temperatures between 12,000~K and 16,000~K. In that cluster we also
found significant disagreement between atmospheric parameters derived
from Str\"omgren photometry or from Balmer line fitting for stars
cooler than about 9,000~K, for which we found no explanation.

Still unexplained are also the low masses found for cool blue HB stars
(which are not affected by diffusion) in, e.g., NGC~6397 and M~92. For
those stars a longer distance scale to globular clusters would reduce
the discrepancies \cite[(Moehler 1999)]{moeh99}.  Such a longer
distance scale has been suggested by several authors using {\sc
Hipparcos} results for metal-poor field subdwarfs to determine the
distances to globular clusters by fitting their main sequence with the
local subdwarfs.  One should note, however, that \cite{detu97}, report
that {\sc Hipparcos} parallaxes for field HBA stars still yield masses
significantly below the canonical mass expected for these objects.
\cite{cagr00} present an extensive and excellent discussion of
various globular cluster distance determinations and the zoo of biases
that affect them, while \cite{grbr03} concentrate on the error budget
of distances from main sequence fitting in their paper.

The problem of the different rotation velocities for cool and hot HB
stars, however, remains rather stubbornly opposed to any attempted solution.

\begin{acknowledgments}
I would like to thank my colleagues and collaborators
Drs. A.V. Sweigart, W.B. Landsman, M. Lemke, U. Heber, and B.B. Behr for many
fruitful discussions and helpful comments. 
\end{acknowledgments}

\begin{discussion}
\discuss{Freytag}{What are ``real'' abundances in contrast to
``diffusion'' abundances? Can they still be reliably defined?}

\discuss{Moehler}{By real abundances I mean the abundances, which the
star would show if its atmosphere were not affected by diffusion (like 
the abundances derived from red giants in globular clusters). And
those cannot be derived if the spectrum {\em is} affected by diffusion}

\discuss{Preston}{Have you compared the abundances of hot HB stars
with those derived from red giants, which are affected in a different
way by diffusion processes?}

\discuss{Moehler}{Red giants should show no diffusion effects in their 
atmospheres due to their deep convection zones. HB stars with
diffusion usually show strong enhancements of heavy elements compared
to red giants in the same cluster.}

\discuss{Corbally}{Some dozen years ago I was working with Richard
Gray on field horizontal branch stars identified by Dave Philip. These 
were all slightly cooler than those you have shown, i.e. truly A-type
horizontal branch stars. We found that they lay near or slightly above 
the main sequence, indeed they were metal-poor as you said, and that
two, from their Balmer line profiles at medium resolution, were
probably helium-rich. Could you comment on the last possibility?}

\discuss{Moehler}{Not really, as I find this a very puzzling result,
which cannot be easily understood in the terms of diffusion.}
\end{discussion} 
\end{document}